\input harvmac
\input epsf

\def\mpade{{\rm Pade}}

\def\frakk#1#2{{{#1}\over{#2}}}

\def \inn{\leftskip = 70 pt\rightskip = 70pt}
\def \out{\leftskip = 0 pt\rightskip = 0pt}
\def\vev#1{\mathopen\langle #1\mathclose\rangle }

\def\sic{supersymmetric}

\def\Acal{{\cal A}}
\def\Bcal{{\cal B}}
\def\Ccal{{\cal C}}
\def\Dcal{{\cal D}}

\def\Acaltilde{{\tilde\Acal}}
\def\Bcaltilde{{\tilde\Bcal}}
\def\Ccaltilde{{\tilde\Ccal}}
\def\Dcaltilde{{\tilde\Dcal}}

\def\npb{{Nucl.\ Phys.\ }{\bf B}}

\def\plb{{Phys.\ Lett.\ }{\bf B}}

\def\sjnp{Sov.\ J.\ Nucl.\ Phys.\ }

{\nopagenumbers
\line{\hfil LTH 396}
\line{\hfil hep-ph/9706249}
\vskip .5in
\centerline{\titlefont Asymptotic Pad\'e Approximants and the 
SQCD $\beta$-function}
\vskip 1in
\centerline{\bf I.~Jack, D.R.T.~Jones }
\medskip
\centerline{\it Dept of Mathematical Sciences, 
University of Liverpool, Liverpool L69 3BX, UK}
\bigskip
\centerline{\bf M.A.~Samuel}
\medskip
\centerline{\it Department of Physics, Oklahoma State University, 
Stillwater, Oklahoma 74078, USA}

\vskip .3in

We present a prediction for the four loop $\beta$-function for SQCD 
based on the method of Asymptotic Pad\' e Approximants. 
\Date{ June 1997}

Recently a four-loop calculation of the gauge $\beta$-function ($\beta_g$)
in an $N=1$ supersymmetric gauge theory  was
presented\ref\jjn{I.~Jack, D.R.T.~Jones and C.G.~North, \npb 486 (1997) 479}.
The computation was performed within the usual supersymmetric dimensional
regularisation scheme (DRED), which is standard for perturbative calculations
in supersymmetry and which consequently would be the most convenient
for any future phenomenological applications. The result was
derived somewhat indirectly, by starting from a
partial calculation for the abelian case. The coupling constant redefinition
relating the abelian result to the corresponding exact NSVZ
result~\ref\nov{
V.~Novikov et al, \npb 229 (1983) 381\semi
V.~Novikov et al, \plb166 (1986) 329\semi
M.~Shifman and A.~Vainstein, \npb 277 (1986) 456\semi
A.~Vainstein, V.~Zakharov and M.~Shifman,
\sjnp 43 (1986) 1028\semi
M.~Shifman, A.~Vainstein and V.~Zakharov \plb 166 (1986) 334} could then
be constructed, and this redefinition was then
extended to the non-abelian  case by imposing the vanishing of the
$\beta$-function beyond one loop in the  case of ${\cal N}=2$
supersymmetry. However, one possible redefinition which was present in
general, but vanished for the abelian case, could not be  determined,
and consequently the four-loop result $\beta_3$  contained one
undetermined constant. Our purpose here is to predict the value of this
constant by using improved Pad\'e approximation techniques which have
recently been developed  and applied to the case of standard 
QCD\ref\EKS{J.~Ellis, M.~Karliner and M.A.~Samuel, hep-ph/9612202}. 
In that case a
prediction  for $\beta_3$ was found to be in good agreement with a  
subsequent  analytic calculation\ref\larin{T.~van Ritbergen,
J.A.M.~Vermaseren and S.A.~Larin, hep-ph/9701390}, 
especially when the contributions
involving  quartic  Casimir group theory structures (which arise for the
first time at  four loops)  are omitted in making the comparison. In
fact these structures  simply do not arise in the \sic\ case which
concerns us here.  This observation, allied with the fact  that we have
more information about $\beta_3$  than was available to Ellis et al in
the QCD case, makes us optimistic  regarding the accuracy of our
prediction. 

The method is as follows: For a perturbative series $P(x)=\sum_{n=0}^{\infty}
S_nx^n$, the Pad\'e approximant $P_{[N/M]}(x)$ is a ratio of polynomials 
$A_N(x)$ and $B_M(x)$, of degree $N$ and $M$ respectively, chosen so that 
\eqn\Pa{
P_{[N/M]}={A_N(x)\over{B_M(x)}}=P(x)+O(x^{N+M+1}). }
It can be argued that in the case of QCD (and presumably also
of \sic\ QCD) the relative error 
\eqn\Pb{\delta_{N+M+1}\equiv
{S^\mpade_{N+M+1} - S_{N+M+1}
\over S_{N+M+1}}}
has the asymptotic form
\eqn\Pc{\delta_{N+M+1}\simeq\,{-} {M! A^M \over L^M} }
as $N\rightarrow\infty$, for fixed $M$, where $A$ is a constant and 
$L=N + M + aM + b$. We will be concerned with $[0,1]$ and $[1,1]$ Pad\'es, so 
$M=1$; and we choose\EKS\ $a$, $b$ so that $a+b=0$. 
The Asymptotic-Pad\'e Approximant Prediction (APAP) is then given by:
\eqn\Pd{
S^{\rm APAP}_{N+M+1} = {S^\mpade_{N+M+1} \over 1 + \delta_{N+M+1}}.}
If $S_0$, $S_1$ and $S_2$ are known, then the 
prediction for $S_3$ is obtained 
as follows. Firstly by matching to $P(x)$ up to linear 
terms as in Eq.~\Pa, we obtain
\eqn\Pe{P_{[0,1]}={S_0^2\over{S_0- S_1 x}},}
giving as a prediction for the coefficient of the $x^2$ term $S_2^{\rm Pade}
={S_1^2\over{S_0}}$. Then from
Eq.~\Pb\ we find $\delta_2={S_1^2\over{S_0S_2}}-1$. From  
Eq.~\Pc\ we deduce $A=-\delta_2$ , and thence   
$\delta_3=-{1\over2}A={1\over2}\delta_2$. The $[1,1]$ Pad\'e prediction for 
$S_3$ is easily derived to be $S_3^{\rm Pade}
={S_2^2\over{S_1}}$, and we then use Eq.~\Pd\ to give an improved estimate
$S_3^{\rm APAP}$ incorporating $\delta_3$. 

The gauge $\beta$-function is given by 
\eqn\Pe{
\beta_g = g\sum_{n=0}^{\infty}\beta_nx^{n+1},}
where $x={g^2\over{16\pi^2}}$, and $\beta_n$ corresponds to the $(n+1)$-loop
result in perturbation theory. The procedure given above yields a prediction for
the four-loop contribution $\beta_3$. The final feature of the method
is to perform the above process for a range of values of $N_f$, the number of
flavours, and then to match the results to a cubic polynomial in $N_f$. 
In the QCD case this enabled the authors of Ref.~\EKS\ to incorporate as 
an extra piece of information the known coefficient of $N_f^3$, obtained by
large-$N_f$ calculations. In our present case we shall be able to exploit the
fact that we know the form of $\beta_3$ as a function of $N_f$ up to a 
single parameter. The results of Ref.~\jjn\ were presented for a general 
theory; for the present purposes, however, we specialise to \sic\ QCD,
obtaining~\ref\jjfn{P.M.~Ferreira, I.~Jack, D.R.T.~Jones and C.G.~North
hep-ph/9705328}:
\eqna\fourb$$\eqalignno{
\beta_0 &= N_f - 3N_c &\fourb a\cr
\beta_1 &= \left[4N_c-{2\over {N_c}}\right]N_f
-6N_c^2 &\fourb b\cr
\beta_2 &= \left[{3\over{N_c}}-4N_c\right]N_f^2
+\left[21N_c^2-{2\over{N_c^2}}-9\right]N_f  
-21N_c^3 &\fourb c\cr
\beta_3 &= \Acal + \Bcal N_f + \Ccal N_f^2 +\Dcal N_f^3 &\fourb d\cr}$$
where $N_c$ is the number of colours, and 
\eqn\loopa{\eqalign{
\Acal &= -(6+36\alpha)N_c^4\cr
\Bcal  &=  36(1+\alpha)N_c^3-(34+12\alpha)N_c-\frakk{8}{N_c}
-\frakk{4}{N_c^3}\cr
\Ccal  &= -\left(\frakk{62}{3}+2\kappa+8\alpha\right)N_c^2+\frakk{100}{3}
+4\alpha+\frakk{6\kappa-20}{3N_c^2}\cr
\Dcal  &= -\frakk{2}{3N_c}.\cr}}
Here $\kappa=6\zeta(3)$ and  
$\alpha$ is the constant which we
hope to determine\foot{It is also possible to apply the APAP method to the 
cases of QED or supersymmetric QED, for which the complete four-loop results
can be extracted from Refs.~\ref\qed{S.G.~Gorishny, A.L.~Kataev, S.A.~Larin and
L.R.~Surguladze, \plb256 (1991) 81}\ or \jjn\ respectively. However, the form 
of the series in these cases is not conducive to the accuracy of the 
approximation and the four-loop predictions are somewhat less impressive. We 
hope to discuss these and other applications of the APAP method elsewhere.}.

Note that $\Dcal $ does not involve $\alpha$. It is the leading-$N_f$ 
contribution at four loops and could, of course, have been 
obtained from the large-$N_f$ results presented in Refs.~\jjfn, 
\ref\jjf{P.M.~Ferreira, I.~Jack, D.R.T.~Jones, hep-ph/9702304}. 
We may therefore proceed in much the same way as did the authors 
of Ref.~\EKS. We first compute $S_3^{\rm APAP}$ as 
described above, and  then after determining $\Acal, \Bcal , \Ccal $ from the 
fit to Eq.~\fourb{d}\ we can obtain three distinct ``predictions'' 
for $\alpha$,  $\alpha(\Acal), \alpha(\Bcal)$ and $\alpha(\Ccal)$. Naturally 
a test of the approach is the extent to which these predictions 
agree and are insensitive to the input value of $N_c$. As in 
Ref.~\EKS\ we use input values $N_f = 0,1,2,3,4$, and used a 
value for $A$, $\vev A$, obtained by averaging $A(N_f)$ for the input values; 
the outcome is not very sensitive, in fact, to whether we use $\vev A$ or 
$A(N_f)$. The results of this procedure are shown in Table 1:

$$\vbox{\offinterlineskip
\def\vr{\vrule height 11pt depth 5pt}
\def\vrq{\vr\quad}
\settabs
\+
\vrq $N_c$  \quad & \vrq  $\alpha(\Acal)$
 \quad& \vrq $\alpha(\Bcal )$\quad&
\vrq $\alpha(\Ccal )$\quad&\vrq $\alpha(\Ccal )\, (\kappa=0)$\quad&\vr\cr\hrule
\+
\vrq $N_c$  \quad & \vrq  $\alpha(\Acal)$
 \quad& \vrq $\alpha(\Bcal)$\quad&
\vrq $\alpha(\Ccal)$\quad&\vrq $\alpha(\Ccal)\, (\kappa=0)$\quad&\vr\cr\hrule
\+
\vrq $2$  \quad & \vrq  $2.56$
 \quad& \vrq $2.34$\quad&
\vrq $-0.31$\quad&\vrq $1.62$\quad&\vr\cr\hrule
\+
\vrq $3$  \quad & \vrq  $2.43$
 \quad& \vrq $2.42$\quad&
\vrq $\phantom{-}0.50$\quad&\vrq $2.38$\quad&\vr\cr\hrule
\+
\vrq $4$  \quad & \vrq  $2.46$
 \quad& \vrq $2.47$\quad&
\vrq $\phantom{-}0.63$\quad&\vrq $2.48$\quad&\vr\cr\hrule
\+
\vrq $5$  \quad & \vrq  $2.45$
 \quad& \vrq $2.46$\quad&
\vrq $\phantom{-}0.74$\quad&\vrq $2.57$\quad&\vr\cr\hrule}$$
\medskip
\inn
{\it \noindent Table~1: APAP predictions for 
the unknown parameter $\alpha$.}
\medskip
\out
The results for $\alpha(\Acal), \alpha(\Bcal)$ are remarkable for 
their consistency and stability. In the fourth column of the table 
we show the result for $\alpha(\Ccal)$ consequent on omitting from 
$\beta_3$ the terms proportional to $\kappa$. 
It could be argued that this is natural since, given that $\zeta(3)$ occurs 
for the first time at four loops, these terms cannot be 
accurately produced by the the Pad\'e. (This is similar to the argument 
made with respect to the quartic Casimir terms in Ref.~\EKS.) In any case, our
results clearly suggest a value for $\alpha$ of around 2.45.

An alternative approach is as follows. The unknown parameter $\alpha$ in 
Eq.~\loopa\ is one of three parameters that were introduced in 
Ref.~\jjn, each 
being the coefficient of a certain coupling constant redefinition.   
Two of these parameters were in fact determined in Ref.~\jjn; 
but we can test the reliability of the APAP method by reintroducing 
one of them and comparing the APAP prediction with a $\beta_3$ now dependent 
on two parameters. Thus we replace $\beta_3$ by $\tilde\beta_3$ where:
\eqn\bdelta{
\beta_3\to \tilde\beta_3 =  
\Acaltilde + \Bcaltilde N_f + \Ccaltilde N_f^2 + \Dcaltilde N_f^3}
and 
\eqn\bdelta{\eqalign{
\Acaltilde &= \Acal,\cr 
 \Bcaltilde &= \Bcal  +\delta\left[24N_c^3 - 12\left(N_c +\frakk{1}{N_c}\right)
\right]\cr 
\Ccaltilde &= \Ccal  + \delta\left[-20N_c^2 +16 +\frakk{4}{N_c^2}\right]\cr
\Dcaltilde &= \Dcal  +\delta\left[4N_c - \frakk{4}{N_c}\right].\cr}}
In the notation of Ref.~\jjn, $\alpha\equiv \alpha_1$ and 
$\delta\equiv -\frakk{2}{3}-\alpha_2$. In Ref.~\jjn\ it was shown that 
$\alpha_2 = -\frakk{2}{3}$, and so we hope to find that $\delta = 0$.  
For fixed $\alpha$ we determine three distinct ``predictions'' for 
$\delta$ by fitting $S_3^{\rm APAP}$ to a polynomial as before. 
As an example, for $N_c = 4$ we obtain:
\eqna\bdeltab$$\eqalignno{
\delta(\Bcaltilde) &= 7.85\alpha-19.24 &\bdeltab a\cr
\delta(\Ccaltilde) &= 19.38\alpha-48.16 &\bdeltab b\cr
\delta(\Dcaltilde) &= 51.94\alpha-126.87. &\bdeltab c\cr}$$
If we omit $\kappa$ terms then Eqs.\bdeltab{a}, \bdeltab{c}\ are 
unaffected, but Eq.\bdeltab{b}\ becomes\hfil\break
$\delta(\Ccaltilde) = 19.38\alpha-47.41$.\hfil  

When we plot $\delta(\Bcaltilde)$, $\delta(\Ccaltilde)$ and
$\delta(\Dcaltilde)$ against $\alpha$, as  in Fig.~1, the results are
quite striking. We see that $\delta(\Bcaltilde)$, $\delta(\Ccaltilde)$
and $\delta(\Dcaltilde)$ are zero for very similar values of $\alpha$,
around 2.44--2.49. So the predicted values for $\delta(\Bcaltilde)$,
$\delta(\Ccaltilde)$ and $\delta(\Dcaltilde)$ agree rather well with the
correct value $\delta=0$ if $2.44\le\alpha\le2.49$, and this may be
interpreted as a prediction for $\alpha$. 

\epsfysize= 3.2in
\centerline{\epsfbox{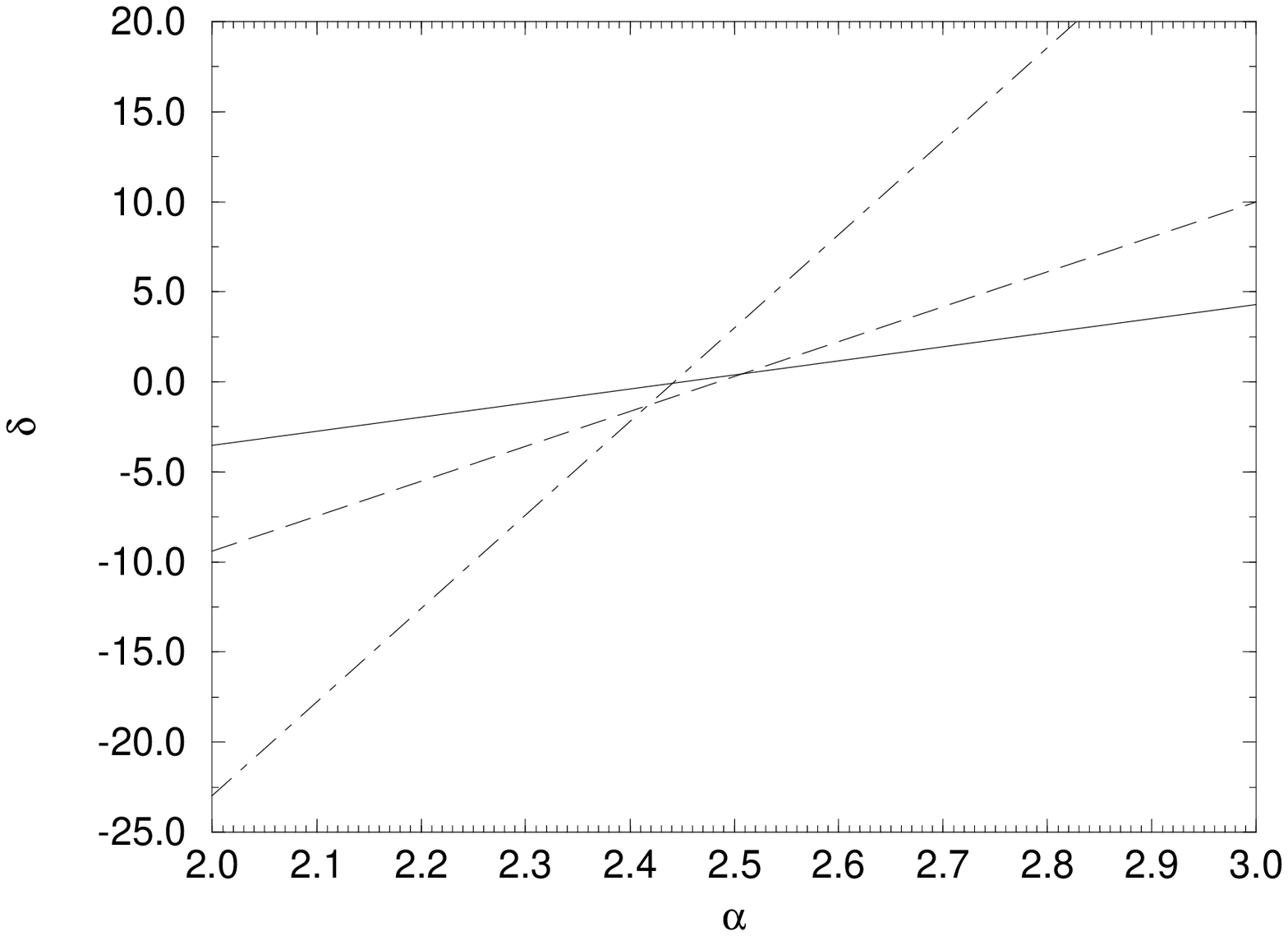}}
\inn
{\it \noindent Fig.~1:
Graph of $\delta(\Bcaltilde)$ (solid line), $\delta(\Ccaltilde)$ (dashed line) 
and $\delta(\Dcaltilde)$ (dash-dotted line) against $\alpha$ for $N_c=4$.}
\medskip
\out
\epsfysize= 3.2in
\centerline{\epsfbox{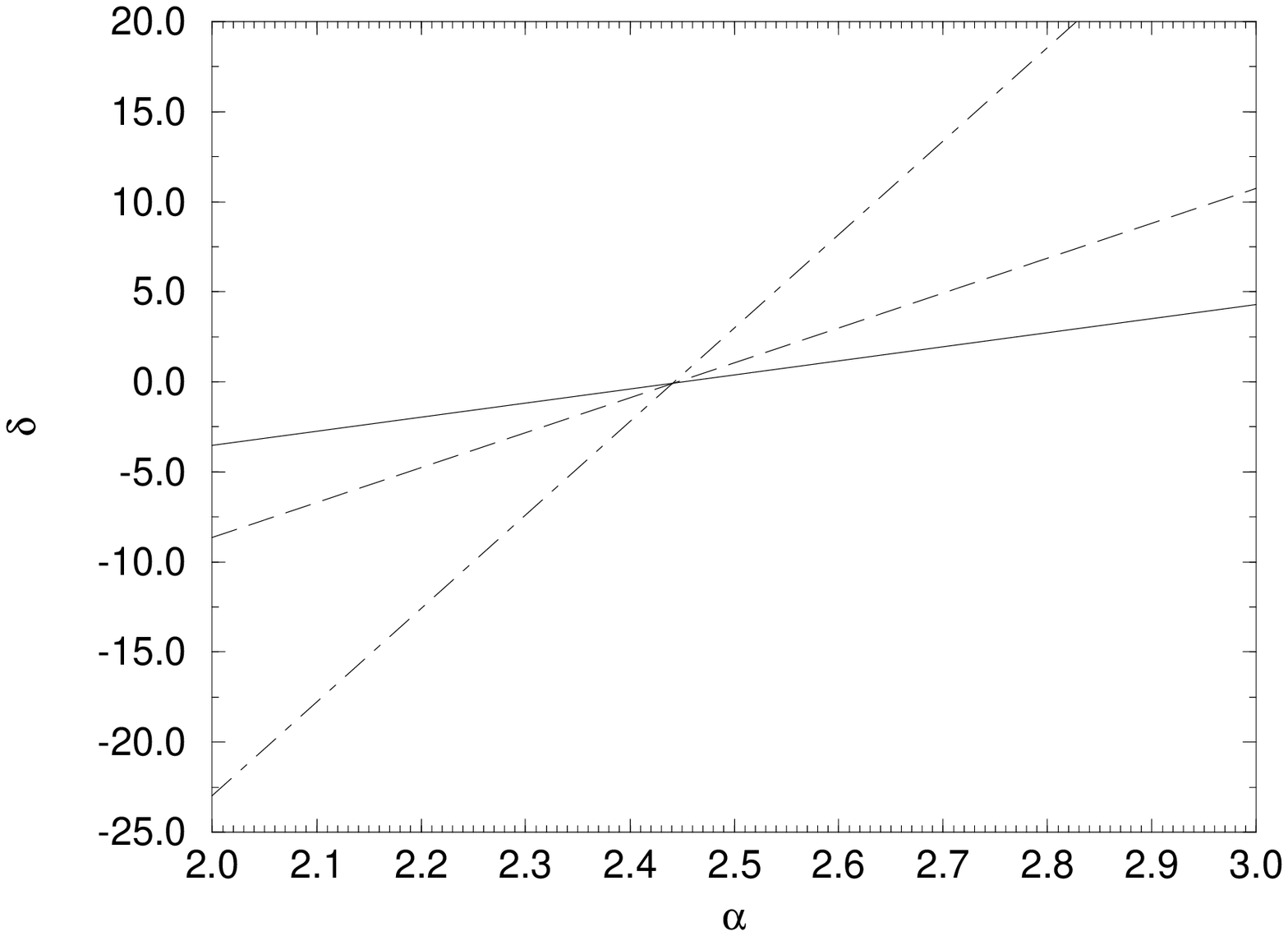}}
\inn
{\it \noindent Fig.~2:
Graph of $\delta(\Bcaltilde)$, $\delta(\Ccaltilde)$  
and $\delta(\Dcaltilde)$
against $\alpha$ for $N_c=4$ (omitting $\kappa$ terms).}
\medskip
\out
The convergence is again even
better if we omit the term in $\kappa$, as depicted in Fig.~2, where
each $\delta$ crosses the axis at $\alpha=2.44$. As $N_c$ is increased,
we find that the convergence of the $\delta$s improves in the case where
we retain $\kappa$; in fact for large $N_c$ all the $\delta$s are zero
for $\alpha=2.43$ irrespective of whether or not we retain $\kappa$.

What is our final prediction for $\alpha$? The relatively simple 
relationship between the NSVZ $\beta$-function and the corresponding DRED 
one as explored in Ref.~\jjn\ suggests that $\alpha$ is a simple fraction; 
given our results here, then we would expect $\alpha = 12/5$ or perhaps 
$\alpha = 5/2$. ($\alpha=17/7$ is even closer, but experience suggests that
this number is unlikely to emerge from a perturbative calculation.)
The result $\alpha = 2\zeta(3)$ is also possible; in 
which case the apparent better convergence produced by excising $\kappa$ 
from  the comparison would need to be dismissed as coincidental. 

\bigskip\centerline{{\bf Acknowledgements}}\nobreak

This work arose out of discussions at the ``Beyond the Standard Model V''
Conference at Balholm, Norway. We thank Per Osland and the other organisers 
of the meeting for providing a pleasant and stimulating atmosphere. 
MAS was supported by the US Department of Energy under Grant No.
DE-FG05-84ER40215.

\listrefs
\bye